\documentclass[seceq]{ptptex}

\usepackage{graphicx}
\usepackage{color}
\usepackage{epsfig}

\renewcommand{\bar}[1]{\overline{#1}}

\renewcommand{\bar}[1]{\overline{#1}}

\newcommand{\ket}[1]{\,\left|\,{#1}\right\rangle}

\def\Dslash{\raise.15ex\hbox{/}\kern-.7em D}
\def\Pslash{\raise.15ex\hbox{/}\kern-.7em P}

\notypesetlogo                       

\title{
The Conformal Template and New Perspectives for Quantum Chromodynamics%
}

\author{
Stanley J. \textsc{Brodsky}%
}

\inst{
Stanford Linear Accelerator Center, Stanford University, Stanford, CA, 94309
}

\abst{
Conformal symmetry provides a  systematic approximation to QCD in
both its perturbative and nonperturbative domains. One can use the
AdS/CFT correspondence between Anti-de Sitter space and conformal
gauge theories to obtain an analytically tractable approximation to
QCD in the regime where the QCD coupling is large and constant. For
example, there is an exact correspondence between the
fifth-dimensional coordinate of AdS space  and a specific impact
variable which measures the separation of the quark constituents
within the hadron in ordinary space-time. This connection allows one
to compute the analytic form of the frame-independent light-front
wavefunctions of mesons and baryons, the fundamental entities which
encode hadron properties and allow the computation of exclusive
scattering amplitudes. One can also use conformal symmetry as a
template for perturbative QCD predictions where the effects of the
nonzero beta function can be systematically included in the scale of
the QCD coupling. This leads to fixing of the renormalization scale
and commensurate scale relations which relate observables without
scale or scheme ambiguity. The results are consistent with the
renormalization group and the analytic connection of QCD to Abelian
theory at $N_C\to 0$. I also discuss a number of novel
phenomenological features of QCD. Initial- and final-state
interactions from gluon-exchange, normally neglected in the parton
model, have a profound effect in QCD hard-scattering reactions,
leading to leading-twist single-spin asymmetries, diffractive deep
inelastic scattering, diffractive hard hadronic reactions, the
breakdown of the Lam Tung relation in Drell-Yan reactions, and
nuclear shadowing and non-universal antishadowing---leading-twist
physics not incorporated in the light-front wavefunctions of the
target computed in isolation.  I also discuss tests of hidden color
in nuclear wavefunctions, the use of diffraction to materialize the
Fock states of a hadronic projectile and test QCD color
transparency, nonperturbative antisymmetric sea quark distributions,
anomalous heavy quark effects, and the unexpected effects of direct
higher-twist processes. }

\begin{document}

\maketitle

\section{Introduction}
This unique conference has illuminated the historical path of the
past 50 years which led to the development of quantum
chromodynamics, starting with Sakata's $p n \Lambda$
proposal~\cite{Sakata:1956}. This model gave the first indication of
$SU(3)$ flavor symmetry and ultimately, a composite theory of
hadrons. The SLAC measurement of inelastic electron-proton
scattering in 1969~\cite{Breidenbach:1969kd} demonstrated the
Bjorken scale invariance~\cite{Bjorken:1968dy} of the deep inelastic
cross section and gave the first indications that the constituents
of the proton are effectively pointlike, thus establishing Gell
Mann~\cite{Gell-Mann:1962xb}, Ne'eman~\cite{Ne'eman:1961cd} and
Zweig's~\cite{Zweig:1964wu} quarks as elementary constituent fields
on par with the leptonic fields of quantum electrodynamics. The
application of Feynman's parton model by Bjorken and
Paschos\cite{Bjorken:1969ja} and Drell, Levy and
Yan~\cite{Drell:1969ca} to deep inelastic lepton-proton scattering
provided the basis for applying quantum field theory  to the strong
interactions. A crucial theoretical development was the advent of
parastatistics by Greenberg~\cite{Greenberg:1964pe} which solved the
apparent contradiction of the quark model with the spin-statistics
problem and provided the basis for $SU(3)$ gauge symmetry. These
historical developments led to the development by Gell-Mann,
Fritzsch and Leutwyler~\cite{Fritzsch:1973pi} of our present  theory
of quantum chromodynamics based on the exact non-Abelian
Yang-Mills~\cite{Yang:1954ek}  $SU(3)$ local color gauge invariance.
The alternative Han-Nambu model~\cite{Han:1965pf} with integral
quark charges was ruled out by hard two-photon
reactions~\cite{Field:1985eb}.

QCD is a fascinating theory with remarkable complexity and novel features.
The physical mechanisms underlying color confinement are still being clarified.
Experiments at RHIC~\cite{Harris:2005sw} are now probing new phenomena associated
with the high temperature  phase of QCD where its
quark and gluon degrees of freedom become explicit.

In this talk I will discuss how conformal symmetry can provide a
systematic approximation to QCD in both its nonperturbative and
perturbative domains. In the case of nonperturbative QCD, one can
use the AdS/CFT correspondence~\cite{Maldacena:1997re} between
Anti-de Sitter space and conformal gauge theories to obtain an
analytically tractable approximation to QCD in the regime where the
QCD coupling is large and constant.  This connection allows one to
compute the analytic form~\cite{Brodsky:2006uq,Brodsky:2007pt} of
the frame-independent light-front wavefunctions of mesons and
baryons, the fundamental entities which encode hadron properties and
allow the computation of exclusive scattering amplitudes.  One can
also use conformal symmetry as a template~\cite{Brodsky:1999gm} for
perturbative QCD expansions where the effects of the nonzero QCD
$\beta$-function can be systematically incorporated into the scale
of the running
coupling~\cite{Brodsky:1994eh,Brodsky:1995tb,Brodsky:2000cr}. This
leads to fixing of the renormalization scale and commensurate scale
relations which relate observables without scale or scheme
ambiguity~\cite{Brodsky:1982gc}. The results are
consistent~\cite{Brodsky:1992pq} with the renormalization
group~\cite{Stueckelberg:1953dz} and the analytic connection  of QCD
to Abelian theory at $N_C\to 0$~\cite{Brodsky:1997jk}.

I will also review in this talk some novel features of QCD. For
example, initial- and final-state interactions normally neglected in
the parton model have a profound effect in QCD hard-scattering
reactions, leading to leading-twist single-spin asymmetries,
diffractive hard hadronic reactions, the breakdown of the Lam Tung
relation in Drell-Yan reactions. Diffractive deep inelastic
scattering leads  the shadowing and antishadowing of nuclear
structure functions---leading-twist physics not incorporated in the
light-front wavefunctions of the target computed in isolation.  I
also will discuss tests of hidden color in nuclear wavefunctions,
the use of diffraction to materialize the Fock states of a hadronic
projectile and test QCD color transparency, nonperturbative
antisymmetric sea quark distributions, anomalous heavy quark
effects, and the unexpected effects of direct higher-twist
processes.  Many of these features of QCD can be tested at RHIC,
$e^+ e^-$ colliders, Fermilab, and the new hadron physics facilities
at JLAB, GSI-FAIR, and J-PARC.

\section{Analytic Connection of QCD to Abelian Theory}
An important guide to perturbative QCD predictions is consistency in
the $N_C \to 0$ limit where the theory becomes
Abelian~\cite{Brodsky:1997jk}. One can consider QCD predictions as
analytic functions of the number of colors $N_C$ and flavors $N_F$.
Remarkably, one can show to all orders of perturbation
theory~\cite{Brodsky:1997jk} that PQCD predictions reduce to those
of an Abelian theory similar to QED at $N_C \to 0$ with $C_F
\alpha_s$ and $N_F/(T_F C_F)$ held fixed, where $C_F=(N^2_C-1)/ (2
N_C)$ and $T_F=1/2.$  The resulting theory corresponds to the group
${1/U(1)}$ -- not $U(1)$  Abelian QED. This means that
light-by-light diagrams acquire a particular topological factor. The
$N_C \to 0$ limit is complimentary to 't Hooft's large $N_C$ limit;
it provides an important check on the analytic behavior of QCD
expressions: QCD formulae and phenomena must match their Abelian
analog. In particular, the renormalization scale in perturbative
expansions is effectively fixed~\cite{Brodsky:1982gc} by this
requirement.

\section{Infrared Fixed Point}

The negative $\beta$ function of the quark-gluon coupling at high
virtuality implies asymptotic freedom and allows the perturbative
analysis of both inclusive and exclusive hard scattering hadronic
reactions. It has usually been assumed that the negative $\beta$ QCD
function also implies ``infrared slavery"; i.e.;  that the QCD
coupling becomes singular in the infrared. However, solutions of the
QCD Dyson Schwinger
equations~\cite{vonSmekal:1997is,Zwanziger:2003cf}  and
phenomenological
studies~\cite{Mattingly:1993ej,Brodsky:2002nb,Baldicchi:2002qm} of
QCD couplings based on physical observables such as $\tau$
decay~\cite{Brodsky:1998ua}  suggest that the QCD $\beta$ function
vanishes and  $\alpha_s(Q^2)$ become constant at small virtuality;
{\em i.e.}, effective charges develop an infrared fixed point.
Recent lattice gauge theory simulations~\cite{Furui:2006py} and
nonperturbative analyses~\cite{Antonov:2007mx} have also indicated
an infrared fixed point for QCD.  One can understand this
physically~\cite{Brodsky:2007pt}: in a confining theory where gluons
have an effective mass or maximal wavelength, all vacuum
polarization corrections to the gluon self-energy decouple at long
wavelength.  When the coupling is constant and quark masses can be
ignored, the  QCD Lagrangian becomes conformally
invariant~\cite{Parisi:1972zy}, allowing the mathematically tools of
conformal symmetry to  be applied, such as the AdS/CFT
correspondence~\cite{Maldacena:1997re}.

\section{Perturbative QCD and Exclusive Processes}

Exclusive processes provide an important window on QCD processes and
the structure of hadrons. Rigorous statements can be made on the
basis of asymptotic freedom and factorization theorems which
separate the underlying hard quark and gluon subprocess amplitude
from the nonperturbative physics of the hadronic wavefunctions. The
leading-power contribution to exclusive hadronic amplitudes such as
quarkonium decay, heavy hadron decay, and scattering amplitudes
where hadrons are scattered with large momentum transfer can often
be factorized as a convolution of distribution amplitudes
$\phi_H(x_i,\Lambda)$ and hard-scattering quark/gluon scattering
amplitudes $T_H$ integrated over the light-front momentum fractions
of the valence quarks~\cite{Lepage:1980fj}: $ {\cal M}_{\rm Hadron}
= \int
 \prod \phi_H^{(\Lambda)} (x_i,\lambda_i)\, T_H^{(\Lambda)} dx_i\ .$
Here $T_H^{(\Lambda)}$ is the underlying quark-gluon subprocess
scattering amplitude in which each incident and final hadron is
replaced by valence quarks with collinear momenta $k^+_i =x_i
p^+_H$, $\vec k_{\perp i} = x_i \vec p_{\perp H }.$ The invariant
mass of all intermediate states in $T_H$ is evaluated above the
separation scale ${\cal M}^2_n > \Lambda^2$. The essential part of
the hadronic wavefunction is the distribution
amplitude~\cite{Lepage:1980fj}, defined as the integral over
transverse momenta of the valence (lowest particle number) Fock
wavefunction.

The leading power fall-off of the hard scattering amplitude as given
by dimensional counting rules follows from the nominal scaling of
the hard-scattering amplitude: $T_H \sim 1/Q^{n-4}$, where $n$ is
the total number of fields (quarks, leptons, or gauge fields)
participating in the hard
scattering~\cite{Brodsky:1974vy,Matveev:1973ra}. Thus the reaction
is dominated by subprocesses and Fock states involving the minimum
number of interacting fields.  In the case of $2 \to 2$ scattering
processes, this implies $ {d\sigma/ dt}(A B \to C D) ={F_{A B \to C
D}(t/s)/ s^{n-2}},$ where $n = N_A + N_B + N_C +N_D$ and $n_H$ is
the minimum number of constituents of $H$. The near-constancy of the
effective QCD coupling at small scales helps explain the empirical
success of dimensional counting rules for the near-conformal power
law fall-off of form factors and fixed angle
scaling~\cite{Brodsky:1989pv}.  For example, one sees the onset of
perturbative QCD scaling behavior even for exclusive nuclear
amplitudes such as deuteron photodisintegration (Here $n = 1+ 6 + 3
+ 3 = 13 .$) $s^{11}{ d\sigma/dt}(\gamma d \to p n) \sim $ constant
at fixed CM angle. The measured deuteron form factor and the
deuteron photodisintegration cross section also appear to follow the
leading-twist QCD predictions at large momentum transfers in the few
GeV region~\cite{Holt:1990ze,Bochna:1998ca,Rossi:2004qm}.

\section{The Conformal Approximation to QCD}

As 't Hooft has emphasized at this meeting, it is important to find
an analytic and tractable first approximation to QCD. In this talk I
will show how conformal symmetry can provide convenient and
systematic approximations to QCD in both its perturbative and
nonperturbative domains.

\subsection{The Conformal Template}  In the case of perturbation theory,
one can use conformal symmetry as a template, systematically
correcting for the nonzero QCD $\beta$
function~\cite{Brodsky:2000cr,Rathsman:2001xe,Grunberg:2001bz}
order-by-order in perturbation theory using the Banks-Zaks
procedure~\cite{Banks:1981nn}. The contributions from the nonzero
$\beta$ function  automatically fix the renormalization scale of the
running QCD coupling consistent with the renormalization group and
the $N_C \to 0$ Abelian limit.   After scale setting, the
perturbative series has the same form as the conformal series and
thus no $n!$ renormalon divergence.  The resulting  predictions for
physical observables are independent of the choice of
renormalization scheme.

The near-conformal behavior of QCD is the basis for commensurate scale
relations~\cite{Brodsky:1994eh} which relate observables to each other without
renormalization scale or scheme ambiguities~\cite{Brodsky:2000cr,Rathsman:2001xe}.  One
can derive the commensurate scale relation between the effective charges of any two
observables by first computing their relation in conformal gauge theory; the effects of
the nonzero QCD $\beta-$ function are then taken into account using the BLM
method~\cite{Brodsky:1982gc} to set the scales of the respective couplings. An important
example is the generalized Crewther relation~\cite{Brodsky:1995tb}:
$
\left[1 + {\alpha_R(s^*)}/{\pi} \right] \left[1 -
{\alpha_{g_1}(Q^2)}/{\pi}\right] = 1
$
where the underlying form at zero $\beta$ function is dictated by conformal
symmetry~\cite{Crewther:1972kn}. Here $\alpha_R(s)/\pi$ and $-\alpha_{g_1}(Q^2)/\pi$
represent the entire radiative corrections to $R_{e^+ e^-}(s)$ and the Bjorken sum rule
for the $g_1(x,Q^2)$ structure function measured in spin-dependent deep inelastic
scattering, respectively. The relation between $s^*$ and $Q^2$ can be computed order by
order in perturbation theory using the BLM method~\cite{Brodsky:1982gc}.   The ratio of
physical scales guarantees that the effect of new quark thresholds is commensurate.
Commensurate scale relations are renormalization-scheme independent and satisfy the group
properties of the renormalization group.  Each observable can be computed in any
convenient renormalization scheme such as dimensional regularization. The $\bar{MS}$
coupling can then be eliminated; it becomes only an intermediary~\cite{Brodsky:1994eh}.

\subsection{Scale-Setting for the Three-Gluon Coupling}
Recently Michael Binger and I ~\cite{Binger:2006sj} have analyzed
the behavior of the 13 nonzero form factors contributing to the
gauge-invariant three-gluon vertex at one-loop, an analysis which is
important for heavy quark production and other PQCD processes.
Supersymmetric relations between scalar, quark, and gluon loops
contributions leads to a simple presentation of the results for a
general non-Abelian gauge theories.  Only the gluon contribution to
the form factors is needed since the massless quark and scalar
contributions are inferred from the homogeneous relation
$F_G+4F_Q+(10-d)F_S=0$ and the sums $\Sigma_{QG}(F) \equiv {(d-2)/
2}F_Q + F_G$ which are given for each form factor $F$. The extension
to the case of internal masses leads to the modified sum rule
$F_{MG}+4F_{MQ}+(9-d)F_{MS}=0$. The phenomenology of the three-gluon
vertex is largely determined by the form factor multiplying the
tree-level tensor.  One can define a three-scale effective scale
$Q^2_{eff}(p^2_a,p^2_b,p^2_c)$ as a function of the three external
virtualities which provides a natural extension of BLM scale setting
\cite{Brodsky:1982gc} to the three-gluon vertex. Physical momentum
scales thus set the scale of the coupling.   The dependence of
$Q^2_{eff}$  on  the physical scales has a number of surprising
features.  A  complicated threshold and pseudo-threshold behavior is
also observed.

In a physical renormalization scheme~\cite{Grunberg:1982fw}, gauge
couplings are defined directly in terms of physical observables.
Such effective charges are analytic functions of the physical scales
and their mass thresholds  have the correct threshold
dependence~\cite{Brodsky:1998mf,Binger:2003by} consistent with
unitarity. As in QED, heavy particles contribute to physical
predictions even at energies below their threshold. This is in
contrast to renormalization schemes such as $\bar{MS}$ where mass
thresholds are treated as step functions.  In the case of
supersymmetric grand unification, one finds a number of qualitative
differences and improvements in precision over conventional
approaches~\cite{Binger:2003by}. The analytic threshold corrections
can be important in making the measured values of the gauge
couplings consistent with unification.

\subsection{AdS/QCD as a First Approximant to Nonperturbative QCD}

\begin{figure}
       \centerline{\includegraphics {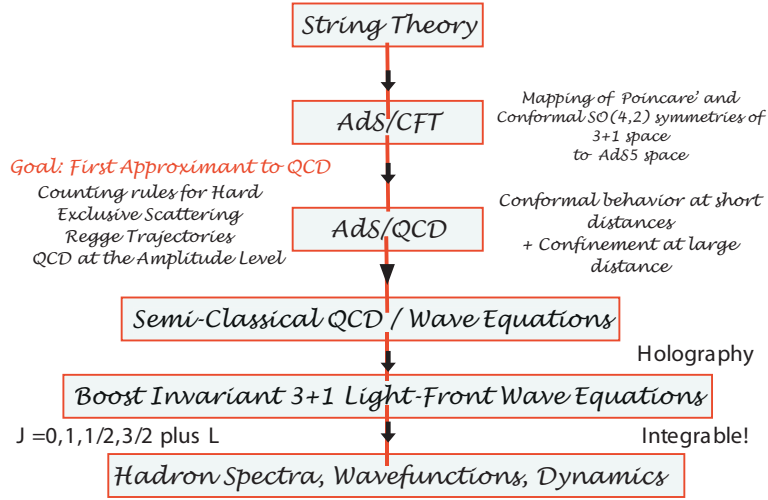}}
   \caption{The AdS/CFT program for QCD.}
   \label{fig:1}
   \end{figure}

The vanishing of the $\beta$ function at small momentum transfer
implies that there is regime where QCD resembles a strongly-coupled
theory and mathematical techniques based on conformal invariance can
be applied. For example, conformal invariance provides the expansion
polynomials for distribution
amplitudes,~\cite{Brodsky:1980ny,Brodsky:1984xk,Brodsky:1985ve,Braun:2003rp,Brodsky:2004qb,Brodsky:2003dn}
the non-perturbative wavefunctions which control exclusive processes
at leading twist~\cite{Lepage:1979zb,Brodsky:2000dr}. One can use
the AdS/CFT correspondence  between Anti-de Sitter space and
conformal gauge theories to obtain an approximation to
nonperturbative QCD in the regime where the QCD coupling is large
and constant;  i.e., one can use the mathematical representation of
the conformal group $S0(4,2)$ in five- dimensional anti-de Sitter
space to construct a holographic representation to the theory. For
example, Guy de Teramond and I ~\cite{Brodsky:2006uq} have shown
that the  amplitude $\Phi(z)$ describing  the hadronic state in the
fifth dimension of Anti-de Sitter space $\rm{AdS}_5$ can be
precisely mapped to the light-front wavefunctions $\psi_{n/h}$ of
hadrons in physical space-time,  thus providing a description of
hadrons in QCD at the amplitude level.  The light-front
wavefunctions are relativistic and frame-independent generalizations
of the familiar Schr\"odinger wavefunctions of atomic physics, but
they are determined at fixed light-cone time $\tau= t +z/c$---the
``front form" advocated by Dirac---rather than at fixed ordinary
time $t$.  We derived this correspondence by noticing that the
mapping of $z \to \zeta$ analytically transforms the expression for
the form factors in AdS/CFT to the exact Drell-Yan-West expression
in terms of light-front wavefunctions. An outline of the application
of AdS/CFT to QCD is shown in Fig.~\ref{fig:1}.

A key result for mesons is an
an effective two-particle light-front radial
equation~\cite{Brodsky:2006uq,Brodsky:2007pt}
\begin{equation}
\label{eq:Scheq} \left[-\frac{d^2}{d \zeta^2} + V(\zeta) \right]
\phi(\zeta) = \mathcal{M}^2 \phi(\zeta),
\end{equation}
with the effective potential $V(\zeta) \to - (1-4 L^2)/4\zeta^2$ the
conformal limit. Here  $\zeta^2 = x(1-x) \mathbf{b}_\perp^2$ where
$x={k^+/ P^+}$ is the light cone momentum fraction. and $b_\perp$ is
the impact separation; i.e. the Fourier conjugate to the relative
transverse momentum $k_\perp$. The variable $\zeta$, $0 \le \zeta
\le \Lambda^{-1}_{\rm QCD}$, represents the invariant separation
between point-like constituents, and it is also the holographic
variable $z$ in AdS; {\em i.e.}, we can identify $\zeta = z$. The
solution to (\ref{eq:Scheq}) is $\phi(z) = z^{-\frac{3}{2}} \Phi(z)
= C z^\frac{1}{2} J_L(z \mathcal{M})$. This equation reproduces the
AdS/CFT solutions. The lowest stable state is determined by the
Breitenlohner-Freedman bound~\cite{Breitenlohner:1982jf}.  We can
model confinement by imposing Dirichlet boundary conditions at
$\phi(z = 1/\Lambda_{\rm QCD}) = 0.$ The eigenvalues are then given
in terms of the roots of the Bessel functions: $\mathcal{M}_{L,k} =
\beta_{L,k} \Lambda_{\rm QCD}$.   Alternatively, one can add a
confinement potential $-\kappa^2 \zeta^2$ to the effective potential
$V(\zeta)$.

With the exception of the pion, the eigenvalues of the effective
light-front equation provide a  good description of the meson and
baryon spectra for light quarks~\cite{Brodsky:2005kc}, and its
eigensolutions provide a remarkably simple but realistic model of
their valence wavefunctions.  The resulting normalized light-front
wavefunctions  for the truncated space model are
\begin{equation}
\widetilde \psi_{L,k}(x, \zeta) =  B_{L,k} \sqrt{x(1-x)} J_L
\left(\zeta \beta_{L,k} \Lambda_{\rm QCD}\right) \theta\big(z \le
\Lambda^{-1}_{\rm QCD}\big),
\end{equation}
where $B_{L,k} = \pi^{-\frac{1}{2}} {\Lambda_{\rm QCD}} \
J_{1+L}(\beta_{L,k})$. The results  display confinement at large inter-quark
separation and conformal symmetry at short distances, thus reproducing
dimensional counting rules for hard exclusive processes.

Given the light-front wavefunctions $\psi_{n/H}(x_i, \vec k_{\perp
i}, \lambda_i )$, one can compute a large range of hadron
observables. For example, the valence, sea-quark and gluon
distributions are defined from the squares of the LFWFS summed over
all Fock states $n$. Form factors, exclusive weak transition
amplitudes~\cite{Brodsky:1998hn} such as $B\to \ell \nu \pi,$ and
the generalized parton distributions~\cite{Brodsky:2000xy} measured
in deeply virtual Compton scattering are (assuming the ``handbag"
approximation) overlaps of the initial and final LFWFS with $n
=n^\prime$ and $n =n^\prime+2$.  The deeply virtual Compton
amplitudes can be Fourier transformed to $b_\perp$ and $\sigma =
x^-P^+/2$ space providing new insights into QCD
distributions~\cite{Burkardt:2005td,Ji:2003ak,Brodsky:2006in,Hoyer:2006xg}.
The distributions in the LF direction $\sigma$ typically display
diffraction patterns arising from the interference of the initial
and final state LFWFs  ~\cite{Brodsky:2006in,Brodsky:2006ku}.  This
can provide a detailed test of  the AdS/CFT LFWFs predictions.

The gauge-invariant distribution
amplitude $\phi_H(x_i,Q)$ defined from the integral over the
transverse momenta $\vec k^2_{\perp i} \le Q^2$ of the valence
(smallest $n$) Fock state provides a fundamental measure of the
hadron at the amplitude level~\cite{Lepage:1979zb,Efremov:1979qk};
they  are the nonperturbative input to the factorized form of hard
exclusive amplitudes and exclusive heavy hadron decays in
perturbative QCD. The resulting distributions obey the DGLAP and
ERBL evolution equations as a function of the maximal invariant
mass, thus providing a physical factorization
scheme~\cite{Lepage:1980fj}. In each case, the derived quantities
satisfy the appropriate operator product expansions, sum rules, and
evolution equations.
It is interesting to note that the distribution amplitude
predicted by AdS/CFT at the hadronic scale
is $\phi_\pi(x, Q _0) = {(4/ \sqrt 3 \pi)}  f_\pi \sqrt{x(1-x)}$
from both the harmonic oscillator and truncated space models is
quite different than the asymptotic distribution amplitude predicted
from the PQCD evolution~\cite{Lepage:1979zb} of the pion distribution
amplitude: $\phi_\pi(x,Q \to \infty)= \sqrt 3  f_\pi x(1-x) $.
The broader shape of the AdS/CFT pion distribution increases the magnitude of
the leading-twist perturbative QCD prediction for the pion form factor
by a factor of $16/9$ compared to the prediction based on the asymptotic form, bringing the
PQCD prediction close to the empirical pion form factor~\cite{Choi:2006ha}.

Hadron form factors can be directly predicted from the overlap
integrals in AdS space or equivalently by using the Drell-Yan-West
formula in physical space-time.   The form factor at high $Q^2$
receives contributions from small $\zeta$, corresponding to small
$\vec b_\perp = {\cal O}(1/Q)$ ( high relative $\vec k_\perp = {\cal
O}(Q)$ as well as $x \to 1$.  The AdS/CFT dynamics is thus distinct
from endpoint models~\cite{Radyushkin:2006iz} in which the LFWF is
evaluated solely at small transverse momentum or large impact
separation.

The $x \to 1$ endpoint domain of structure functions is often
referred to as a "soft" Feynman contribution. In fact $x \to 1$ for
the struck quark requires that all of the spectators have $x =
k^+/P^+ = (k^0+ k^z)/P^+  \to 0$; this in turn requires high
longitudinal momenta $k^z \to - \infty$ for all spectators  --
unless one has both massless spectator quarks $m \equiv 0$ with zero
transverse momentum $k_\perp \equiv 0$, which is a regime of measure
zero. If one uses a covariant formalism, such as the Bethe-Salpeter
theory, then the virtuality of the struck quark  becomes  infinitely
spacelike:  $k^2_F \sim  - {(k^2_\perp + m^2)/(1-x)}$  in the
endpoint domain. Thus, actually,  $x \to 1$ corresponds to high
relative longitudinal momentum; it is as hard a domain in the hadron
wavefunction as high transverse momentum. Note also that  at large
$x$ where the struck quark is far-off shell, DGLAP evolution is
quenched~\cite{Brodsky:1979qm}, so that the fall-off of the DIS
cross sections in $Q^2$ satisfies inclusive-exclusive duality at
fixed $W^2.$

\section{Higher Fock States}
Since they are complete and orthonormal, the AdS/CFT model
wavefunctions can also be used as a basis for the diagonalization of
the full light-front QCD Hamiltonian, thus systematically improving
the AdS/CFT approximation.   In particular this procedure could provide the higher Fock states
of the QCD hadronic eigenstates.

The physics of
higher Fock states such as the $\vert uud q \bar Q \rangle$
fluctuation of the proton is clearly nontrivial; the phenomenological
distributions display asymmetric
$\bar u(x) \ne \bar d(x)$, and $s(x) \ne \bar s(x)$ sea quark distributions,  and
intrinsic heavy quarks $c \bar c$ and $b \bar b$ which have their
support at high momentum~\cite{Brodsky:2000sk} . Color adds an extra
element of complexity: for example there are five-different color
singlet combinations of six $3_C$ quark representations which appear
in the deuteron's valence wavefunction, leading to the hidden-color
phenomena~\cite{Brodsky:1983vf}.

\subsection{The Strange Quark Asymmetry}
In the simplest treatment of deep inelastic scattering, nonvalence
quarks are produced via gluon splitting and DGLAP evolution.
However, in the full theory, heavy quarks are multiply connected to
the valence quarks~\cite{Brodsky:1980pb}. Although the strange and
antistrange distributions in the nucleon are identical when they
derive from gluon-splitting $g \to s \bar s$, this is not the case
when the strange quarks are part of the intrinsic structure of the
nucleon -- the multiple interactions of the sea quarks produce an
asymmetry of the strange and anti-strange distributions in the
nucleon due to their different interactions with the other quark
constituents.  A QED analogy is the distribution of $\tau^+$ and
$\tau^-$ in a higher Fock state of muonium $\mu^+ e^-.$  The
$\tau^-$ is attracted to the higher momentum $\mu^+$ thus
asymmetrically distorting its momentum distribution. Similar effects
will  happen in QCD. If we use the diquark model $\ket p  \sim
\ket{u_{3_c} (ud)_{\bar 3_C}},$ then the $Q_{3_C}$ in the $\ket{u
(ud) Q \bar Q }$ Fock state will be attracted to the heavy diquark
and thus have higher rapidity than the $\bar Q$. An alternative
model is the $\ket{K \Lambda}$ fluctuation model for the $\ket{uud s
\bar s}$ Fock state of the proton~\cite{Brodsky:1996hc}. The $s$
quark tends to have higher $x$.

Empirical evidence  continues to accumulate that the
strange-antistrange quark distributions are not symmetric in the
proton~\cite{Brodsky:1996hc,Kretzer:2004bg,Portheault:2004xy}. The
experimentally observed asymmetry appears to be small but positive:
$\int dx x [s(x)- \bar s(x)] > 0.$ The results of a recent CTEQ
global data analysis of neutrino-induced dimuon
data are given in ref.  \cite{Olness:2003wz} .
The shape of the strangeness
asymmetry is consistent with the $\Lambda K$ fluctuation
model~\cite{Brodsky:1996hc}. Kretzner~\cite{Kretzer:2004bg} has
noted that a significant part of the NuTeV anomaly could be due to
this asymmetry, The $\bar s(x)-s(x)$ asymmetry can be studied in
detail in $p \bar p$ collisions by searching for antisymmetric
forward-backward strange quark distributions in the $\bar p-p$ CM
frame.

\subsection{Intrinsic Heavy Quarks}
The probability for Fock states of a light hadron such as the proton
to have an extra heavy quark pair decreases as $1/m^2_Q$ in
non-Abelian gauge theory~\cite{Franz:2000ee,Brodsky:1984nx}.  The
relevant matrix element is the cube of the QCD field strength
$G^3_{\mu \nu}.$  This is in contrast to abelian gauge theory where
the relevant operator is $F^4_{\mu \nu}$ and the probability of
intrinsic heavy leptons in QED bound state is suppressed as
$1/m^4_\ell.$  The intrinsic Fock state probability is maximized at
minimal off-shellness.  It is useful to define the transverse mass
$m_{\perp i}= \sqrt{k^2_{\perp i} + m^2_i}.$ The maximum probability
then occurs at $x_i = { m^i_\perp /\sum^n_{j = 1} m^j_\perp}$; {\em
i.e.}, when the constituents have minimal invariant mass and equal
rapidity. Thus the heaviest constituents have the highest momentum
fractions and the highest $x_i$. Intrinsic charm thus predicts that
the charm structure function has support at large $x_{bj}$ in excess
of DGLAP extrapolations~\cite{Brodsky:1980pb}; this is in agreement
with the EMC measurements~\cite{Harris:1995jx}.
Intrinsic charm can also explain the $J/\psi \to \rho \pi$
puzzle~\cite{Brodsky:1997fj}. It also affects the extraction of
suppressed CKM matrix elements in $B$ decays~\cite{Brodsky:2001yt}.

The
dissociation of the intrinsic charm $|uud c \bar c>$
Fock state of the proton on a nucleus can produce a leading heavy
quarkonium state at high $x_F = x_c + x_{\bar c}~$ in $p A \to
J/\psi X A^\prime$ since the $c$ and $\bar c$ can readily coalesce
into the charmonium state.  Since the constituents of a given
intrinsic heavy-quark Fock state tend to have the same rapidity,
coalescence of multiple partons from the projectile Fock state into
charmed hadrons and mesons is also favored.  For example,
one can produce a leading $\Lambda_c$
at high $x_F$ and low $p_T$ from the coalescence of the $u d c$
constituents of the projectile $|uud c \bar c>$  Fock state.  A similar coalescence
mechanism was used in atomic physics to produce relativistic
antihydrogen in $\bar p A$ collisions~\cite{Munger:1993kq}. This
phenomena is important not only for understanding heavy-hadron
phenomenology, but also for understanding the sources of neutrinos
in astrophysics experiments~\cite{Halzen:2004bn}
and the ``long-flying" component in cosmic rays~\cite{Dremin:2005dm}.

In the case of a nuclear target, the charmonium state will be
produced at small transverse momentum and high $x_F$  with a
characteristic $A^{2/3}$ nuclear dependence since the color-octet
color-octet $|(uud)_{8C} (c \bar c)_{8C} >$ Fock state interacts on
the front surface of the nuclear target~\cite{Brodsky:2006wb}. This
forward contribution is in addition to the $A^1$ contribution
derived from the usual perturbative QCD fusion contribution at small
$x_F.$   Because of these two components, the cross section violates
perturbative QCD factorization for hard inclusive
reactions~\cite{Hoyer:1990us}.  This is consistent with the observed
two-component cross section for charmonium production observed by
the NA3 collaboration at CERN~\cite{Badier:1981ci} and more recent
experiments~\cite{Leitch:1999ea}. The diffractive dissociation of
the intrinsic charm Fock state leads to leading charm hadron
production and fast charmonium production in agreement with
measurements~\cite{Anjos:2001jr}.  Intrinsic charm can also explain
the $J/\psi \to \rho \pi$ puzzle~\cite{Brodsky:1997fj},  and it
affects the extraction of suppressed CKM matrix elements in $B$
decays~\cite{Brodsky:2001yt}.

The production cross section for the double-charm $\Xi_{cc}^+$
baryon~\cite{Ocherashvili:2004hi} and the production of $J/\psi$
pairs appears to be consistent with the diffractive dissociation and
coalescence of double IC Fock states~\cite{Vogt:1995tf}.  It is
unlikely that the appearance of two heavy quarks at high $x_F$ could
be explained by the ``color drag model" used in PYTHIA
simulations~\cite{Andersson:1983ia} in which the heavy quarks are
accelerated from low to high $x$ by the fast valence quarks. These
observations provide compelling evidence for the diffractive
dissociation of complex off-shell Fock states of the projectile and
contradict the traditional view that sea quarks and gluons are
always produced perturbatively via DGLAP evolution. It is also
conceivable that the observations~\cite{Bari:1991ty} of $\Lambda_b$
at high $x_F$ at the ISR in high energy $p p$  collisions could be
due to the diffractive dissociation and coalescence of the
``intrinsic bottom" $|uud b \bar b>$ Fock states of the proton.

Intrinsic heavy quarks can also enhance the production probability of Higgs
bosons at hadron colliders from processes such as $g c \to H c.$ It
is thus critical for new experiments (HERMES, HERA, COMPASS) to
definitively establish the phenomenology of the charm structure
function at large $x_{bj}.$
Recently Kopeliovich, Schmidt, Soffer, and I ~\cite{Brodsky:2006wb}
have  proposed a novel mechanism for exclusive diffractive
Higgs production $pp \to p H p $  in which the Higgs boson
carries a significant fraction of the projectile proton momentum.
The production mechanism is based on the subprocess $(Q \bar Q) g \to H $
where the $Q \bar Q$ in the $|uud Q \bar Q>$ intrinsic heavy quark Fock
state has up to $80\%$ of the projectile protons momentum.
This process will provide a clear experimental signal for
Higgs production due to the small background in this kinematic region.

\subsection{Hidden Color}
In traditional nuclear physics, the deuteron is a bound state of a
proton and a neutron where the binding force arise from the exchange
of a pion and other mesonic states. However, QCD provides a new
perspective:~\cite{Brodsky:1976rz,Matveev:1977xt}  six quarks in the
fundamental $3_C$ representation of $SU(3)$ color can combine into
five different color-singlet combinations, only one of which
corresponds to a proton and neutron.  In fact, if the deuteron
wavefunction is a proton-neutron bound state at large distances,
then as their separation becomes smaller, the QCD evolution
resulting from colored gluon exchange introduce four other ``hidden
color" states into the deuteron wavefunction~\cite{Brodsky:1983vf}.
The normalization of the deuteron form factor observed at large
$Q^2$~\cite{Arnold:1975dd}, as well as the presence of two mass
scales in the scaling behavior of the reduced deuteron form
factor,~\cite{Brodsky:1976rz} thus suggest sizable hidden-color Fock
state contributions such as $\ket{(uud)_{8_C} (ddu)_{8_C}}$ with
probability  of order $15\%$ in the deuteron
wavefunction~\cite{Farrar:1991qi}.

The hidden color states of the deuteron can be materialized at the
hadron level as $\Delta^{++}(uuu) \Delta^{-}(ddd)$ and other novel
quantum fluctuations of the deuteron.  These dual hadron components
become more and more important as one probes the deuteron at short
distances, such as in exclusive reactions at large momentum
transfer.  For example, the ratio ${{d \sigma/ dt}(\gamma d \to
\Delta^{++} \Delta^{-})/{d\sigma/dt}(\gamma d\to n p) }$ should
increase dramatically with increasing transverse momentum $p_T.$
Similarly the Coulomb dissociation of the deuteron into various
exclusive channels $e d \to e^\prime + p n, p p \pi^-, \Delta
\Delta, \cdots$ should have a changing composition as the
final-state hadrons are probed at high transverse momentum,
reflecting the onset of hidden color degrees of freedom.

\section{Diffractive Deep Inelastic Scattering}
A remarkable feature of deep inelastic lepton-proton scattering at
HERA is that approximately 10\% events are
diffractive~\cite{Adloff:1997sc,Breitweg:1998gc}: the target proton
remains intact, and there is a large rapidity gap between the proton
and the other hadrons in the final state.  These diffractive deep
inelastic scattering (DDIS) events can be understood most simply
from the perspective of the color-dipole model: the $q \bar q$ Fock
state of the high-energy virtual photon diffractively dissociates
into a diffractive dijet system.  The exchange of multiple gluons
between  the color dipole of the $q \bar q$ and the quarks of the
target proton neutralizes the color separation and leads to the
diffractive final state.  The same multiple gluon exchange also
controls diffractive vector meson electroproduction at large photon
virtuality.\cite{Brodsky:1994kf}  This observation presents a
paradox: if one chooses the conventional parton model frame where
the photon light-front momentum is negative $q+ = q^0 + q^z  < 0$,
the virtual photon interacts with a quark constituent with
light-cone momentum fraction $x = {k^+/p^+} = x_{bj}.$  Furthermore,
the gauge link associated with the struck quark (the  Wilson line)
becomes unity in light-cone gauge $A^+=0$. Thus the struck
``current" quark apparently experiences no final-state interactions.
Since the light-front wavefunctions $\psi_n(x_i,k_{\perp i})$ of a
stable hadron are real, it appears impossible to generate the
required imaginary phase associated with pomeron exchange, let alone
large rapidity gaps.

This paradox was resolved by Hoyer, Marchal,  Peigne, Sannino and
myself.\cite{Brodsky:2002ue} Consider the case where the virtual
photon interacts with a strange quark---the $s \bar s$ pair is
assumed to be produced in the target by gluon splitting.  In the
case of Feynman gauge, the struck $s$ quark continues to interact in
the final state via gluon exchange as described by the Wilson line.
The final-state interactions occur at a light-cone time $\Delta\tau
\simeq 1/\nu$ shortly after the virtual photon interacts with the
struck quark. When one integrates over the nearly-on-shell
intermediate state, the amplitude acquires an imaginary part. Thus
the rescattering of the quark produces a separated color-singlet $s
\bar s$ and an imaginary phase. In the case of the light-cone gauge
$A^+ = \eta \cdot A =0$, one must also consider the final-state
interactions of the (unstruck) $\bar s$ quark. The gluon propagator
in light-cone gauge $d_{LC}^{\mu\nu}(k) = (i/k^2+ i
\epsilon)\left[-g^{\mu\nu}+\left(\eta^\mu k^\nu+ k^\mu\eta^\nu /
\eta\cdot k\right)\right] $ is singular at $k^+ = \eta\cdot k = 0.$
The momentum of the exchanged gluon $k^+$ is of ${ \cal
O}{(1/\nu)}$; thus rescattering contributes at leading twist even in
light-cone gauge.   The net result is  gauge invariant and is
identical to the color dipole model calculation. The calculation of
the rescattering effects on DIS in Feynman and light-cone gauge
through three loops is given in detail for an Abelian model in the
references.~\cite{Brodsky:2002ue}    The result shows that the
rescattering corrections reduce the magnitude of the DIS cross
section in analogy to nuclear shadowing.

A new understanding of the role of final-state interactions in deep
inelastic scattering has thus emerged. The multiple  scattering of
the struck parton via instantaneous interactions in the target
generates dominantly imaginary diffractive amplitudes, giving rise
to an effective ``hard pomeron" exchange.  The presence of a
rapidity gap between the target and diffractive system requires that
the target remnant emerges in a color-singlet state; this is made
possible in any gauge by the soft rescattering.  The resulting
diffractive contributions leave the target intact  and do not
resolve its quark structure; thus there are contributions to the DIS
structure functions which cannot be interpreted as parton
probabilities~\cite{Brodsky:2002ue}; the leading-twist contribution
to DIS  from rescattering of a quark in the target is a coherent
effect which is not included in the light-front wave functions
computed in isolation. One can augment the light-front wave
functions with a gauge link corresponding to an external field
created by the virtual photon $q \bar q$ pair
current.\cite{Belitsky:2002sm,Collins:2004nx}   Such a gauge link is
process dependent~\cite{Collins:2002kn}, so the resulting augmented
LFWFs are not
universal.\cite{Brodsky:2002ue,Belitsky:2002sm,Collins:2003fm}   We
also note that the shadowing of nuclear structure functions is due
to the destructive interference between multi-nucleon amplitudes
involving diffractive DIS and on-shell intermediate states with a
complex phase. In contrast, the wave function of a stable target is
strictly real since it does not have on-energy-shell intermediate
state configurations.   The physics of rescattering and shadowing is
thus not included in the nuclear light-front wave functions, and a
probabilistic interpretation of the nuclear DIS cross section is
precluded.

Rikard Enberg, Paul Hoyer, Gunnar Ingelman and
I~\cite{Brodsky:2004hi} have shown that the quark structure function
of the effective hard pomeron has the same form as the quark
contribution of the gluon structure function. The hard pomeron is
not an intrinsic part of the proton; rather it must be considered as
a dynamical effect of the lepton-proton interaction. Our QCD-based
picture also applies to diffraction in hadron-initiated processes.
The rescattering is different in virtual photon- and hadron-induced
processes due to the different color environment, which accounts for
the  observed non-universality of diffractive parton distributions.
This framework also provides a theoretical basis for the
phenomenologically successful Soft Color Interaction (SCI)
model~\cite{Edin:1995gi} which includes rescattering effects and
thus generates a variety of final states with rapidity gaps.

\section{ Single-Spin Asymmetries from Final-State
Interactions}
Among the most interesting polarization effects are single-spin
azimuthal asymmetries  in semi-inclusive deep inelastic scattering,
representing the correlation of the spin of the proton target and
the virtual photon to hadron production plane: $\vec S_p \cdot \vec
q \times \vec p_H$.  Such asymmetries are time-reversal odd, but
they can arise in QCD through phase differences in different spin
amplitudes. In fact, final-state interactions from gluon exchange
between the outgoing quarks and the target spectator system lead to
single-spin asymmetries in semi-inclusive deep inelastic
lepton-proton scattering  which  are not power-law suppressed at
large photon virtuality $Q^2$ at fixed
$x_{bj}.$~\cite{Brodsky:2002cx}   In contrast to the SSAs arising from
transversity and the Collins fragmentation function, the
fragmentation of the quark into hadrons is not necessary; one
predicts a correlation with the production plane of the quark jet
itself. Physically, the final-state interaction phase arises as the
infrared-finite difference of QCD Coulomb phases for hadron wave
functions with differing orbital angular momentum.  The same proton
matrix element which determines the spin-orbit correlation $\vec S
\cdot \vec L$  also produces the anomalous magnetic moment of the
proton, the Pauli form factor, and the generalized parton
distribution $E$ which is measured in deeply virtual Compton
scattering. Thus the contribution of each quark current to the SSA
is proportional to the contribution $\kappa_{q/p}$ of that quark to
the proton target's anomalous magnetic moment $\kappa_p = \sum_q e_q
\kappa_{q/p}$.\cite{Brodsky:2002cx,Burkardt:2004vm}  The HERMES
collaboration has recently measured the SSA in pion
electroproduction using transverse target
polarization.\cite{Airapetian:2004tw}  The Sivers and Collins
effects can be separated using planar correlations; both
contributions are observed to contribute, with values not in
disagreement with theory
expectations.\cite{Airapetian:2004tw,Avakian:2004qt} A related
analysis also predicts that the initial-state interactions from
gluon exchange between the incoming quark and the target spectator
system lead to leading-twist single-spin asymmetries in the
Drell-Yan process $H_1 H_2^\updownarrow \to \ell^+ \ell^- X$.
\cite{Collins:2002kn,Brodsky:2002rv}    The SSA in the Drell-Yan
process is the same as that obtained in SIDIS, with the appropriate
identification of variables, but with the opposite sign.
There is no Sivers effect in charged-current reactions since the $W$
only couples to left-handed quarks.\cite{Brodsky:2002pr}

If both the quark and antiquark in the initial state of the Drell-Yan subprocess
$q \bar q \mu^+ \mu^-$ interact with the spectators of the other incident hadron,
one finds a breakdown of the Lam-Tung relation,  which was formerly believed
to be a general prediction of leading-twist QCD.
These double initial-state interactions also lead to a $\cos 2 \phi$ planar
correlation in unpolarized Drell-Yan reactions.\cite{Boer:2002ju}
More generally one must consider subprocesses involving initial-state gluons
such as $n g q \bar q \to \ell \bar \ell$  in addition to subprocesses with extra final-state gluons.

The final-state interaction mechanism provides an appealing physical
explanation within QCD of single-spin asymmetries.  Remarkably, the
same matrix element which determines the spin-orbit correlation
$\vec S \cdot \vec L$  also produces the anomalous magnetic moment
of the proton, the Pauli form factor, and the generalized parton
distribution $E$ which is measured in deeply virtual Compton
scattering.  Physically, the final-state interaction phase arises as
the infrared-finite difference of QCD Coulomb phases for hadron wave
functions with differing orbital angular momentum.  An elegant
discussion of the Sivers effect including its sign has been given by
Burkardt~\cite{Burkardt:2004vm}.
As shown recently by Gardner and myself~\cite{Brodsky:2006ha}, one can also use
the Sivers effect to study the orbital angular momentum of  gluons
by tagging a gluon jet in semi-inclusive DIS. In this case,
the final-state interactions are enhanced by the large color charge of the gluons.

The final-state interaction effects can also be identified with the
gauge link which is present in the gauge-invariant definition of
parton distributions~\cite{Collins:2004nx}.  Even when the light-cone gauge
is chosen, a transverse gauge link is required.  Thus in any gauge
the parton amplitudes need to be augmented by an additional eikonal
factor incorporating the final-state interaction and its
phase~\cite{Ji:2002aa,Belitsky:2002sm}. The net effect is that it is
possible to define transverse momentum dependent parton distribution
functions which contain the effect of the QCD final-state
interactions.

\section{Diffraction Dissociation as a Tool to Resolve Hadron
Substructure and Test Color Transparency}
Diffractive multi-jet production in heavy nuclei provides a novel
way to resolve the shape of light-front Fock state wave functions
and test color transparency.\cite{Brodsky:1988xz}   For example,
consider the reaction~\cite{Bertsch:1981py,Frankfurt:1999tq} $\pi A
\rightarrow {\rm Jet}_1 + {\rm Jet}_2 + A^\prime$ at high energy
where the nucleus $A^\prime$ is left intact in its ground state. The
transverse momenta of the jets balance so that $ \vec k_{\perp i} +
\vec k_{\perp 2} = \vec q_\perp < {R^{-1}}_A \ . $ Because of color
transparency, the valence wave function of the pion with small
impact separation will penetrate the nucleus with minimal
interactions, diffracting into jet pairs.\cite{Bertsch:1981py}  The
$x_1=x$, $x_2=1-x$ dependence of the dijet distributions will thus
reflect the shape of the pion valence light-cone wave function in
$x$; similarly, the $\vec k_{\perp 1}- \vec k_{\perp 2}$ relative
transverse momenta of the jets gives key information on the second
transverse momentum derivative of the underlying shape of the
valence pion wavefunction.\cite{Frankfurt:1999tq,Nikolaev:2000sh}
The diffractive nuclear amplitude extrapolated to $t = 0$ should be
linear in nuclear number $A$ if color transparency is correct.  The
integrated diffractive rate will then scale as $A^2/R^2_A \sim
A^{4/3}.$ This is in fact what has been observed by the E791
collaboration at FermiLab for 500 GeV incident pions on nuclear
targets~\cite{Aitala:2000hc}.  The measured momentum fraction
distribution of the jets with high transverse momentum is found to be approximately consistent
with the shape of the pion asymptotic distribution amplitude,
$\phi^{\rm asympt}_\pi (x) = \sqrt 3 f_\pi
x(1-x)$~\cite{Aitala:2000hb}; however, there is an indication
from the data that the distribution is broader at lower transverse momentum,
consistent with the AdS/CFT prediction.

Color transparency, as evidenced by the Fermilab measurements of
diffractive dijet production, implies that a pion can interact
coherently throughout a nucleus with minimal absorption, in dramatic
contrast to traditional Glauber theory based on a fixed $\sigma_{\pi
n}$ cross section.  Color transparency gives direct validation of
the gauge interactions of QCD.
Color transparency has also been observed in diffractive
electroproduction of $\rho$ mesons \cite{Borisov:2002rd} and in
quasi-elastic $p A \to p p (A-1)$ scattering~\cite{Aclander:2004zm}
where only the small size fluctuations of the hadron wavefunction
enters the hard exclusive scattering amplitude.  In the latter case
an anomaly occurs at $\sqrt s \simeq 5 $ GeV, most likely signaling
a resonance effect at the charm threshold~\cite{Brodsky:1987xw}.

\section{Shadowing and Antishadowing of Nuclear Structure Functions}
One of the novel features of QCD involving nuclei is the {\it
antishadowing} of the nuclear structure functions which is observed
in deep inelastic lepton scattering and other hard processes.
Empirically, one finds $R_A(x,Q^2) \equiv  \left(F_{2A}(x,Q^2)/
(A/2) F_{d}(x,Q^2)\right) > 1 $ in the domain $0.1 < x < 0.2$; {\em
i.e.}, the measured nuclear structure function (referenced to the
deuteron) is larger than than the scattering on a set of $A$
independent nucleons.

The shadowing of the nuclear structure
functions: $R_A(x,Q^2) < 1 $ at small $x < 0.1 $ can be readily
understood in terms of the Gribov-Glauber theory.  Consider a
two-step process in the nuclear target rest
frame. The incoming $q \bar q$ dipole first interacts diffractively
$\gamma^* N_1 \to (q \bar q) N_1$ on nucleon $N_1$ leaving it
intact.  This is the leading-twist diffractive deep inelastic
scattering  (DDIS) process which has been measured at HERA to
constitute approximately 10\% of the DIS cross section at high
energies.  The $q \bar q$ state then interacts inelastically on a
downstream nucleon $N_2:$ $(q \bar q) N_2 \to X$. The phase of the
pomeron-dominated DDIS amplitude is close to imaginary, and the
Glauber cut provides another phase $i$, so that the two-step process
has opposite  phase and  destructively interferes with the one-step
DIS process $\gamma* N_2 \to X$ where $N_1$ acts as an unscattered
spectator. The one-step and-two step amplitudes can coherently
interfere as long as the momentum transfer to the nucleon $N_1$ is
sufficiently small that it remains in the nuclear target;  {\em
i.e.}, the Ioffe length~\cite{Ioffe:1969kf} $L_I = { 2 M \nu/ Q^2} $
is large compared to the inter-nucleon separation. In effect, the
flux reaching the interior nucleons is diminished, thus reducing the
number of effective nucleons and $R_A(x,Q^2) < 1.$

There are also leading-twist diffractive contributions $\gamma^* N_1
\to (q \bar q) N_1$  arising from Reggeon exchanges in the
$t$-channel.\cite{Brodsky:1989qz}  For example, isospin--non-singlet
$C=+$ Reggeons contribute to the difference of proton and neutron
structure functions, giving the characteristic Kuti-Weisskopf
$F_{2p} - F_{2n} \sim x^{1-\alpha_R(0)} \sim x^{0.5}$ behavior at
small $x$. The $x$ dependence of the structure functions reflects
the Regge behavior $\nu^{\alpha_R(0)} $ of the virtual Compton
amplitude at fixed $Q^2$ and $t=0.$ The phase of the diffractive
amplitude is determined by analyticity and crossing to be
proportional to $-1+ i$ for $\alpha_R=0.5,$ which together with the
phase from the Glauber cut, leads to {\it constructive} interference
of the diffractive and nondiffractive multi-step nuclear amplitudes.
Furthermore, because of its $x$ dependence, the nuclear structure
function is enhanced precisely in the domain $0.1 < x <0.2$ where
antishadowing is empirically observed.  The strength of the Reggeon
amplitudes is fixed by the fits to the nucleon structure functions,
so there is little model dependence.

As noted above, the Bjorken-scaling diffractive contribution
to DIS arises from the rescattering of the struck quark after it is
struck  (in the  parton model frame $q^+ \le 0$), an effect induced
by the Wilson line connecting the currents. Thus one cannot
attribute DDIS to the physics of the target nucleon computed in
isolation.\cite{Brodsky:2002ue}  Similarly, since shadowing and
antishadowing arise from the physics of diffraction, we cannot
attribute these phenomena to the structure of the nucleus itself:
shadowing and antishadowing arise because of the $\gamma^* A$
collision and the history of the $q \bar q$ dipole as it propagates
through the nucleus.

Ivan Schmidt, Jian-Jun Yang, and
I~\cite{Brodsky:2004qa} have extended the Glauber analysis to the shadowing
and antishadowing of all of the electroweak structure functions.
Quarks of different flavors  will couple to different Reggeons; this
leads to the remarkable prediction that nuclear antishadowing is not
universal; it depends on the quantum numbers of the struck quark.
This picture leads to substantially different antishadowing for
charged and neutral current reactions, thus affecting the extraction
of the weak-mixing angle $\theta_W$.  We find that part
of the anomalous NuTeV result~\cite{Zeller:2001hh} for $\theta_W$
could be due to the non-universality of nuclear antishadowing for
charged and neutral currents. Detailed measurements of the nuclear
dependence of individual quark structure functions are thus needed
to establish the distinctive phenomenology of shadowing and
antishadowing and to make the NuTeV results definitive. Schmidt,
Yang, and I have also identified contributions to the nuclear
multi-step reactions which arise from odderon exchange and
hidden color degrees of freedom in the nuclear wavefunction. There
are other ways in which this new view of antishadowing can be
tested;  antishadowing can also depend on the target and beam
polarization.

\section{Higher Twist Effects}
Although the contributions of higher twist processes are suppressed
at high transverse momentum, there are some areas of phenomenology
where they can play a dominant role. For example, hadrons can
interact directly within a hard subprocess, leading to higher twist
contributions which can actually dominate over leading twist
processes~\cite{Berger:1979du,Berger:1981fr}.  A classic example is
the reaction $\pi q \to \ell^+ \ell^- q^\prime$ which dominates
Drell-Yan reactions $\pi N \to \ell^+ \ell^- X$ at high $x_F$ and
produces longitudinally polarized lepton pairs.

Higher-twist reactions~\cite{Brodsky:2005fz} such as $ u u \to p
\bar d$ and $(uud) u \to p u$ can dominate single inclusive hadron
reactions at high transverse momentum such as $ p p \to p X$ at high
$x_T = 2p_T/\sqrt s$.  Such ``direct" reactions can explain the
fast-falling power-law  falloffs observed at fixed $x_T$ and fixed
$\theta_{cm}$ observed at the ISR, FermiLab and RHIC.  A review of
the fixed $x_T$ scaling data is given in ref.~\cite{Brodsky:2005fz}.

\section*{Acknowledgments}
Presented at the International Symposium, ``The Jubilee Of The
Sakata Model ($p n \Lambda 50$)", November 25-26, 2006, in Nagoya,
Japan. I thank the organizers for their invitation to speak at this
meeting. I also thank my collaborators for many helpful discussions.
This work was supported in part by the Department of Energy,
contract No. DE-AC02-76SF00515.

\end{document}